\documentclass{ws-ijmpd}
\usepackage[super,compress]{cite}
\begin{document}
\title{Scalar perturbation potentials in a homogeneous and isotropic Weitzenb\"{o}ck geometry}
\author{A. Behboodi}
\address{Department of Physics, Faculty of Basic Sciences,
University of Mazandaran, P. O. Box 47416-95447\\ Babolsar,
IRAN\\a.behboodi@stu.umz.ac.ir}

\author{S. Akhshabi}%
\address{ Department of Physics, Golestan University, P. O.
Box 49138-15759\\ Gorgan, IRAN,\\ s.akhshabi@gu.ac.ir}
\author{K. Nozari}
\address{Department of Physics, Faculty of Basic Sciences,
University of Mazandaran, P. O. Box 47416-95447\\ Babolsar,
IRAN\\k.nozari@umz.ac.ir}

\maketitle \vspace{2cm}
\begin{abstract}
We describe the fully gauge invariant cosmological perturbation
equations in teleparallel gravity by using the gauge covariant
version of the Stewart lemma for obtaining the variations in tetrad
perturbations. In teleparallel theory, perturbations are the result
of small fluctuations in the tetrad field. The tetrad transforms as
a vector in both its holonomic and anholonomic indices. As a result,
in the gauge invariant formalism, physical degrees of freedom are
those combinations of perturbation parameters which remain invariant
under a diffeomorphism in the coordinate frame, followed by an
arbitrary rotation of the local inertial (Lorentz) frame. We derive
these gauge invariant perturbation potentials for scalar
perturbations and present the gauge invariant field equations
governing their evolution.
\end{abstract}
 \ccode{PACS numbers: 04.50.Kd}

\section{Introduction}

Standard cosmology, based on the assumption of isotropy and
homogeneity at large scales, has remarkable success in describing
the known properties of the universe. The origin of local
inhomogeneous structures such as galaxies and galaxy clusters then
can be traced to small fluctuations around the FRW background of
standard cosmology. Generally in cosmology, computing the
perturbation of a quantity involves determining the difference
between the value of this quantity in the actual physical spacetime
and the value of it in the reference unperturbed FRW frame. To
determine this difference, it is necessary to compare the values at
the same spacetime point in both frames. To do this, one needs to
choose a map that shows which point on the perturbed spacetime is
the same as a given point in the background geometry. Choosing a map
is called a gauge choice and therefore changing the map implies
performing a gauge transformation. Generally the choice of the map
is arbitrary. This is referred to as the gauge freedom of the
perturbation theory. There are two different approaches in
performing calculations in this setup. One method is to choose a
gauge (or identification map) and perform the calculation in this
gauge. This method, while it's usually simpler, has a drawback. It
may result in the so called 'gauge mode' solutions which are not
real physical modes. The second method, first introduced by Bardeen
in \cite{Baar82}, involves finding some gauge invariant combinations
of perturbation modes and rewriting the equations in terms of this
gauge invariant quantities. This approach has the advantage of
having only real unambiguous physical quantities but it is
technically more involved. The gauge invariant theory of
cosmological perturbations in general relativity, has been developed
through the years and applied to various problems including
inflation theory and computing the spectrum of the cosmic microwave
background radiation. For thorough reviews of this method see
references \cite{KS84,mfb92,Durrer89,Riotto}.

Teleparallel theory of gravity first introduced by Einstein
\cite{Ein30} in an attempt to unify gravity with electromagnetism.
This theory can be regarded as the gauge theory for the translation
group and in its general form does not possess the same symmetrical
structure as the theory of general relativity (GR) \cite{Hehl07}.
Translational gauge theory is formulated in terms of coframe: in
each point of a n-dimensional manifold one introduces n linearly
independent vector as the basis of spacetime and their dual
covectors or coframes. The theory then possess a torsion field which
can be regarded as the translational field strength corresponding to
the coframe field \cite{Haya79}. The coframe theory in its general
form, does only possess the diffeomorphism invariance and the
invariance under the global Lorentz transformation of tetrads. It is
not invariant under a local Lorentz transformation. Restoring the
local Lorentz invariance means imposing some restrictions on the
form of the Lagrangian. Doing this will result in a theory which is
dynamically equivalent to general relativity and is usually called
the teleparallel equivalent of general relativity (TEGR) in the
literature \cite{Ald10,TG}. Teleparallel gravity and its extensions
\cite{LI,FT} have generated renewed interest in recent years, in the
hope that it may offer solutions to some of the cosmological
problems like the origin and the nature of the dark energy
\cite{NO1,NO2}. It should be stated here that both teleparallel
gravity and general relativity can be regarded as special cases of a
more fundamental gauge theory of gravity called Poincare gauge
theory (PGT) which contains both torsion and curvature as
translational and rotational field strengths respectively
\cite{Blag}. Teleparallel gravity could not achieve the unification
of forces desired by Einstein, nonetheless there are significant
incentives in studying this theory and its properties. For example,
according to some authors a quantization approach based on the
teleparallel variables will probably appear much more natural and
consistent compared to general relativity \cite{Ald10}. Also it has
been shown that owing to its gauge structure, attempts to
incorporate the nonlocal effects into the theory of gravity can best
be done in a teleparallel formulation. \cite{MH}.

For the reasons stated above, it seems interesting to study the
cosmological consequences of teleparallel theories of gravity. One
of the differences between GR and TEGR is in defining the dynamical
variables of the theory. In TEGR unlike GR tetrad fields play the
important role of dynamical variables. This fact has motivated the
authors to consider the cosmological perturbations by directly
perturbing the tetrad fields and study the potential differences. In
this paper we also attempt to obtain fully gauge invariant
formulation of equations governing perturbation dynamics.

\section{Notation and definitions}

Throughout the paper, the Greek indices $\mu,\nu,..$ run over
$0,1,2,3$ and refer to the spacetime coordinates; The Greek letters
from the beginning of the alphabet $\alpha,\beta,...$ run over
$1,2,3$ and refer to spatial coordinates, middle Latin letters
$i,j,...$ run over $0,1,2,3$ and refer to the 4D tangent space
coordinates and finally Latin letters from the beginning of the
alphabet $a,b,...$ run over $1,2,3$ and indicate the spatial tangent
space coordinates.

As stated, in teleparallel gravity one considers a set of n linearly
independent vectors  $ e_{i}=e^{\mu}_{~i}\partial_{\mu}$ which form
a basis in the tangent space on every point of the manifold. The
dual of this basis $\vartheta^{i}=e_{~\mu}^{i}dx^{\mu}$ are
coframes. The dynamical variable in TEGR ($e_{~\mu}^{i}$) are called
the tetrads and they relate anholonomic tangent space indices to the
coordinate ones. The spacetime metric is not an independent
dynamical variable here and is related to the tetrad through the
relations

\begin{equation}
g_{\mu\nu}=\eta_{ij}e_{\mu}^{i}e_{\nu}^{j},
\end{equation}
\begin{equation}
\textbf{e}_{i}\,.\,\textbf{e}_{j}=\eta_{ij}
\end{equation}
The inverse of the tetrad is  defined by the relation
$e_{i}^{\mu}e_{\mu}^{j}=\delta_{i}^{j}.$

Teleparallel geometry, $T_{4}$ is obtained by the requirement of
vanishing curvature. In the special case of TEGR the spin connection
of the theory is also assumed to be zero. This assumption is usually
called the absolute parallelism condition and by imposing it the
connection of the theory  will be Weitzenb\"{o}ck connection defined
as \cite{Weitz23}
\begin{equation}
\Gamma^{\rho}_{\,\,\,\,\mu\nu}=e^{\rho}_{i}\partial_{\nu}e^{i}_{\mu}
\end{equation}
which unlike Livi-civita connection is not symmetric on its second
and third indices. The curvature of this connection is identically
zero and the torsion tensor is
\begin{equation}
T^{\rho}_{\,\,\,\,\mu\nu}\equiv
e_{i}^{\rho}(\partial_{\mu}e_{\nu}^{i}-\partial_{\nu}e_{\mu}^{i})\,.
\end{equation}
Contorsion tensor which denotes the difference between Livi-civita
and Weitzenb\"{o}ck connections is
\begin{equation}
K^{\mu\nu}_{\quad\rho}=-\frac{1}{2}(T^{\mu\nu}_{\quad\rho}
-T^{\nu\mu}_{\quad\rho}-T_{\rho}^{\,\,\,\,\mu\nu})
\end{equation}
and the superpotential tensor is defined as
\begin{equation}
S^{\,\,\,\,\mu\nu}_{\rho}=\frac{1}{2}(K^{\mu\nu}_{\quad\rho}
+\delta^{\mu}_{\rho}T^{\alpha\nu}_{\quad\alpha}-\delta^{\nu}_{\rho}
T^{\alpha\mu}_{\quad\alpha})\,.
\end{equation}
In correspondence with Ricci scalar, one can define torsion scalar
\begin{equation}
T=S^{\,\,\,\,\mu\nu}_{\rho}T^{\rho}_{\,\,\,\,\mu\nu}
\end{equation}
The gravitational action in TEGR is
\begin{equation}
I=\frac{1}{16\pi G}\int dx\,|e|\, T
\end{equation}
where $|e|$ is the determinant of $e^{a}_{\mu}$ and from the
relation (1), one can easily finds it equal to $\sqrt{-g}$.
Variation of the above action with respect to tetrads gives the
field equations of TEGR
\begin{equation}
e^{-1}\partial_{\mu}(ee_{i}^{\rho}S^{\,\,\,\,\mu\nu}_{\rho})
-e_{i}^{\lambda}T_{\,\,\,\,\mu\lambda}^{\rho}S^{\,\,\,\,\nu\mu}_{\rho}
+\frac{1}{4}e_{i}^{\nu}T=4\pi G e_{i}^{\rho}\Xi^{\,\,\nu}_{\rho}
\end{equation}
where $\Xi^{\,\,\nu}_{\rho}$ is the energy-momentum tensor.\\
\section{Gauge invariant cosmological perturbations in teleparallel
gravity}

We now begin the process of deriving fully gauge invariant
cosmological perturbation equations in teleparallel gravity. For a
given quantity $Q$, its perturbation $\delta Q$ is the difference
between its value in the actual perturbed spacetime and in the
background reference geometry, calculated at the same point
\begin{equation}
\delta Q(p)=Q(p)-Q^{\,0}(D^{-1}(p))
\end{equation}
where $D$ is the diffeomorphism $D: N\rightarrow M$ from the
background manifold $N$ to the physical manifold $M$. Another
diffeomorphism, say $\tilde{D}$, result in a different
$\tilde{\delta Q(p)}$
\begin{equation}
\tilde{\delta Q(p)}=\tilde{Q(p)}-Q^{\,0}(\tilde{D}^{-1}(p))
\end{equation}

The change of the identification map between the two manifolds
$D\rightarrow \tilde{D}$ results in a gauge transformation $\delta
Q(p)\rightarrow \tilde{\delta Q(p)}$. As another viewpoint, a gauge
change can also be regarded as a coordinate transformation on the
background manifold
\begin{equation}
x^{\mu}\rightarrow \tilde{x}^{\mu}=x^{\mu}+\xi^{\mu}
\end{equation}
This transformation will lead to a change in $\delta Q$ as
\begin{equation}
\Delta \delta Q=\tilde{\delta Q}-\delta Q=L_{\xi} Q
\end{equation}
where $L$ is the Lie derivative in the direction of $\xi$. A
quantity then called gauge invariant if
\begin{equation}
\Delta \delta Q=L_{\xi} Q=0
\end{equation}
This important result was first derived in \cite{Stewart90} (see
also \cite{SW}) and is usually called the 'Stewart Lemma' in the
literature.

As mentioned before teleparallel theory in its general form can be
regarded as the gauge theory for the translation group. In such a
gauge theory, the notion of ordinary Lie derivative used in Stewart
lemma in general relativity, will no longer be adequate. Here, the
variation of a field is obtained by using the gauge covariant Lie
derivative with respect to a vector $v$, defined for a Lie-algebra
valued form $\Psi$ as \cite{MAG,Nester1,Nester2}
\begin{equation}
L_{v}^{(c)}\Psi=v\rfloor D\Psi +D(v\rfloor\Psi)
\end{equation}
where $\rfloor$ is the interior product and $D$ is the covariant
derivative, here obtained by using the Weitzenbock connection (7).
In coordinate notation, using the gauge covariant form of the
stewart lemma, the variation in the perturbations of the tetrad
field will be
\begin{equation}
\Delta \delta e_{~~\mu}^{i}=e_{\nu}^{i}\nabla_{\mu} \xi^{\nu}
\end{equation}

\section{Tetrad perturbations and gauge invariant potentials}
We begin by considering the unperturbed Friedmann-Robertson-Walker
line element using a conformal time parameter
\begin{equation}
ds^2=a^2(\tau)\{-d\tau^{2}+\gamma_{\alpha\beta}dx^{\alpha}x^{\beta}\}
\end{equation}
where $\gamma_{\alpha\beta}$ is the metric of the spacelike surfaces
of constant torsion in teleparallel gravity and we denotes the
covariant derivative associated with it by $D_{\alpha}$. This metric
will be used to raise and lower indices in the spatial hypersurface.
for simplicity we assume
$\gamma_{\alpha}^{\beta}=\delta_{\alpha}^{\beta}$ where
$\delta_{\alpha}^{\beta}$ is the Kronecker delta. \\
The simplest tetrad describing this FRW background is given by
\begin{equation}
\quad\quad\quad\quad\quad\quad e^{i}_{\mu}= \left(
\begin{array}{cccc}
a & 0 & 0 & 0  \nonumber\\
0 & a & 0 &0
 \nonumber\\
0& 0 & a&0\nonumber\\
0& 0 & 0&a\nonumber
\end{array}
\right) \,
\end{equation}
As we are working in a TEGR setup, any other possible FRW tetrad is
related to this by a Lorentz transformation and will result in the
same dynamics.

The tetrad transforms as a vector in the coordinate space so
generally like any other vector $A^{\alpha}$, they can be decomposed
to a scalar part and a purely vector part according to
\begin{equation}
A^{\alpha}=D^{\alpha}\phi+B^{\alpha}
\end{equation}
where $\phi$ is a scalar field and $B^{\alpha}$ is a solenoidal
vector i.e. $D_{\alpha}B^{\alpha}=0$. The general perturbed FRW
tetrad of teleparallel gravity is given by
\begin{eqnarray}
\nonumber e_{0}^{\bar{0}}&=&a(1+A) \quad , \quad
e_{\alpha}^{\bar{0}}=a\,C_{\alpha} \quad , \quad
e_{0}^{a}=a\,B^{a},\\
 e_{\alpha}^{a}&=&a\,(\delta_{\alpha}^{a}+D_{\alpha}^{a})
\end{eqnarray}
and its inverse
\begin{eqnarray}
\nonumber h_{\bar{0}}^{0}&=&\frac{1}{a}(1-A) \quad , \quad
h^{\alpha}_{\bar{0}}=-\frac{1}{a}\,B^{\alpha} \quad , \quad
h^{0}_{a}=-\frac{1}{a}\,C_{i},\\
h^{\alpha}_{a}&=&\frac{1}{a}\,(\delta^{\alpha}_{a}-D^{\alpha}_{a})
\end{eqnarray}
where an over-dot denotes a tangent space index. Here $A$ is a
scalar and $B^{i}$ , $C_{\alpha}$ and $D_{\alpha}^{i}$ act as vector
degrees of freedom. The tensor part of the perturbations in this
formalism, comes from the metric as can bee seen in the appendix
where $D_{\alpha\beta}=\eta_{ij}\delta_{\alpha}^{i}D_{\beta}^{j}$
contains the tensor part of the perturbations. Note that
$D_{\alpha\beta}$ is still of the first order in tetrad
perturbations.

 If we
decompose $\xi^{\mu}$ in (12) into temporal and spatial parts as
\begin{equation}
\xi^{\mu}=(T, L^{\alpha})
\end{equation}
then using stewart lemma (14) and using the covariant form of the
Lie derivative defined in (15), under transformation (12) we have
\begin{eqnarray}
\Delta A=T'+h T \quad , \quad \Delta B^{i}=L'^{i}+hL^{i}\\ \nonumber
\Delta C_{\alpha}=\partial_{\alpha}T \quad\quad , \quad\quad \Delta
D_{\alpha}^{i}= D_{\alpha}L^{i}
\end{eqnarray}
here a prime denotes differentiation with respect to proper time and
$h=a'/a$. For scalar perturbations, we should have
\begin{equation}
B^{i}=D^{i}B \quad , \quad  C_{\alpha}=D_{\alpha}C  \quad , \quad
D_{\alpha}^{\alpha}=D
\end{equation}
the scalar parameters transform as
 \begin{eqnarray}
A\rightarrow A+T'+h T \\ \nonumber B\rightarrow B+L'+h L \\
\nonumber C\rightarrow C+T \\ \nonumber D\rightarrow D+L
 \end{eqnarray}
where $\Delta=D_{\alpha}D^{\alpha}$ is the spatial Laplacian. The
above perturbation parameters are not gauge invariant; however there
exist two different combination of them which are gauge invariant
\begin{eqnarray}
\Phi_{1}&\equiv& A-C'-h C\\ \nonumber \Phi_{2}&\equiv&B-D'-hD
\end{eqnarray}
These two parameters are gauge invariant potentials and corresponds
to Bardeen's potentials of general relativity. It is obvious that
they are not equivalent. By rewriting the perturbed field equations
in terms of these parameters, one can deal with only real physical
quantities and any gauge ambiguities will be removed. The reason for
different gauge invariant potential in TEGR and GR is twofold. One
reason is that in teleparallel gravity, one should use the
teleparallel version of the Lie derivative in the Stewart lemma
(14), defined in a weitzenb\"{o}ck $T_{4}$ geometry. The other
source of difference is the fact that unlike metric in general
relativity, tetrads does not have any tensor modes.
\section{Matter perturbations}
We assume an unperturbed energy-momentum tensor in the form of a
perfect fluid given by
\begin{equation}
\Xi_{\mu\nu}=(\rho+p)u_{\mu}u_{\nu}+p g_{\mu\nu}
\end{equation}
where $u_{\mu}$ is the 4-velocity. Under the gauge transformation we
have
\begin{eqnarray}
\delta \Xi_{\alpha\beta}\rightarrow \delta \Xi_{\alpha\beta}+T p a^2
\hat{p} \delta_{\alpha\beta} +2pa^2D_{(\alpha}L_{\beta)}\\ \nonumber
\delta \Xi_{00}\rightarrow \delta \Xi_{00}+\rho a^2
[\hat{\rho}T+2T'+2hT]\\ \nonumber \delta \Xi_{0\alpha}\rightarrow
\delta \Xi_{0\alpha}+a^2[p L'_{\alpha}+p h L_{\alpha}+\rho
D_{\alpha}T]
\end{eqnarray}
where $\hat{\rho}=\frac{\rho'}{\rho}$ and $\hat{p}=\frac{p' }{p}$.
The perturbation in the energy-momentum tensor can be written as
\begin{equation}
\delta \Xi^{\mu\nu}=\rho \delta\rho u^{\mu}u^{\nu}+2a
q^{(\mu}u^{\nu)}+p \delta p(u^{\mu}u^{\nu}+g^{\mu\nu})+p a^2
\pi^{\mu\nu}
\end{equation}
where $q^{\mu}$ is the momentum perturbation and $\pi^{\mu\nu}$ is
the anisotropic stress. Under gauge transformations we have
\begin{eqnarray}
\delta \rho\rightarrow \delta \rho+\hat{\rho}T+2T'+2hT \nonumber\\
q_{\alpha} \rightarrow q_{\alpha}-p L'_{\alpha}-\rho D_{\alpha}T-p h
L_{\alpha}\nonumber\\ \delta p\rightarrow \delta
p+\hat{p}T+\frac{2}{3}D_{\alpha}L^{\alpha}\nonumber\\
\pi_{\alpha\beta} \rightarrow
\pi_{\alpha\beta}+2D_{(\alpha}L_{\beta)}-\frac{2}{3}\delta_{\alpha\beta}D_{\gamma}L^{\gamma}
\end{eqnarray}
For scalar perturbations $L_{\alpha}=D_{\alpha}L$,
$q_{\alpha}=D_{\alpha}q$ and
$\pi_{\alpha\beta}=\Delta_{\alpha\beta}\pi$, so
\begin{eqnarray}
\delta \rho\rightarrow \delta \rho+\hat{\rho}T+2T'+2hT  \nonumber\\
q \rightarrow q-p L'-\rho T-p h L\nonumber\\
\nonumber \delta p\rightarrow \delta p+\hat{p}T+\frac{2}{3}\Delta
L\nonumber\\ \pi \rightarrow \pi+2L
\end{eqnarray}

There are four different gauge invariant scalar combinations of
matter and tetrad perturbation variables
\begin{eqnarray}
\Phi_{\rho}=\delta \rho-2A-\hat{\rho} C \nonumber \\ \nonumber
\Phi_{q}=q+\rho C+p B \\ \nonumber \Phi_{p}=\delta p-\hat{p}C-\frac{2}{3}\Delta D\\
\Phi_{\pi}=\pi-2D
\end{eqnarray}
In the next section we will present various components of the
teleparallel field equation (9) in terms of the six gauge invariant
variable presented here.
\section{Gauge invariant field equations}
The torsion, contorsion and superpotential tensor of the perturbed
tetrad (19) is given in the appendix. Various components of the
teleparallel field equation are
\begin{eqnarray}
(i=0,\nu=0) :
&~&\frac{1}{a^{4}}\Bigg[\partial_{\alpha}\Big(\frac{h}{a}B^{\alpha}-a\partial^{[\alpha}D_{~~\beta]}^{\beta}
-\frac{3}{2}h\,a(C^{\alpha}-B^{\alpha})\Big)\Bigg]\nonumber\\
&+&\frac{h}{a^{3}}(\frac{5}{2}+\frac{1}{a^{2}})(D_{~~\alpha}^{'\alpha}-\partial_{\alpha}B^{\alpha})
+\frac{9}{2}\frac{h^{2}}{a^{5}} (1-3A)=4\pi
G e_{.0}^{\alpha} \Xi^{0}_{\alpha}\nonumber\\
\end{eqnarray}
\begin{eqnarray}
(i=a,\nu=0)
:&-&\frac{1}{2a^{3}}\partial_{\alpha}\Big[-\partial_{(a}B^{\alpha)}+\delta^{\alpha}_{a}\partial_{\beta}B^{\beta}
+\partial^{[\alpha}C_{a]}
+D_{~~a)}^{'(\alpha}-\delta_{a}^{\alpha}D_{~~\beta}^{'\beta}\nonumber\\
&+&\frac{2h}{a^{2}}\delta_{a}^{\alpha}(A-|D|)+\frac{2h}{a^{2}}D_{~~a}^{\alpha}\Big]-\frac{2h}{a^{5}}\partial_{[\beta}D_{~~a]}^{\beta}
-\frac{9}{2}\frac{h^{2}}{a^{5}}C_{a}\nonumber\\
&=&4\pi G e_{a}^{\alpha} \Xi^{0}_{\alpha}
\end{eqnarray}

\begin{eqnarray}
(i=a,\nu=\alpha)&:&\frac{1}{2a^{3}}\partial_{\beta}\Big[-\partial
^{[\alpha}D_{~a]}^{\beta}+\partial^{[\beta}D_{~a]}^{\alpha}+\partial^{[\beta}D_{a}^{~\alpha]}
+2\delta^{~\beta}_{a}\partial^{[\alpha}D_{~\eta]}^{\eta}-2\delta^{~\alpha}_{a}\partial^{[\beta}D_{~\eta]}^{\eta}\Big]\nonumber\\
&+&\frac{1}{a^{4}}\partial_{0}\Bigg[\frac{-a}{2}\Big(-\partial_{(a}B^{\alpha)}+\partial^{\alpha}_{a}\partial_{\beta}B^{\beta}
+\frac{2h}{a^{2}}D_{a}^{\alpha}
-\delta_{a}^{\alpha}D_{~~\beta}^{'\beta}
+D_{~~a)}^{'(\alpha}+\partial^{[\alpha}C_{a]}\nonumber\\
&-&\frac{2h}{a^{2}}(1-A-|D|)\delta^{\alpha}_{a}\Big)\Bigg]
-\frac{1}{a^{2}}(A+|D|)\partial_{0}
(\frac{h}{a^{3}}\delta^{\alpha}_{a})+\frac{h}{2a^{3}}\Big[\partial_{(a}B^{\alpha)}\nonumber\\
&-&\frac{2}{a^{2}}\partial_{a}B^{\alpha}\Big]
-\frac{2h}{a^{3}}\delta^{\alpha}_{a}\partial_{\beta}B^{\beta}
-\frac{h}{2a^{3}}\Big[D_{~~a)}^{'(\alpha}-\frac{2}{a^{2}}D_{~~a}^{'\alpha}\Big]+\frac{2h}{a^{3}}D_{\beta}^{'\beta}\delta^{\alpha}_{a}\nonumber\\
&+&\frac{5}{2}\frac{h^{2}}{a^{2}}\delta^{\alpha}_{a}(1-2A)
-\frac{h}{2a^{3}}\partial^{[\alpha}C_{a]}
-\frac{5}{2}\frac{h^{2}}{a^{5}}D_{~~a}^{\alpha}=4\pi G
e_{a}^{\beta}\Xi^{\alpha}_{\beta}
\end{eqnarray}
\begin{eqnarray}
(i=0,\nu=\alpha)&:&\frac{1}{2a^{3}}\partial_{\beta}\Big[D^{'[\beta\alpha]}-\partial^{[\alpha}B^{\beta]}-\partial^{[\beta}C^{\alpha]}\Big]\nonumber\\
&+&\frac{1}{a^{4}}\partial_{0}\Big[\frac{-h}{a}B^{\alpha}+a\partial^{[\alpha}D_{~~\beta]}^{\beta}+\frac{3}{2}h\,a(C^{\alpha}-B^{\alpha})\Big]\nonumber\\
&-&\frac{h}{2a}\Big[3\partial^{[\alpha}D_{~~\beta]}^{\beta}+\partial^{[\beta}D^{~~\alpha]}_{\beta}\Big]
-\frac{5}{2}\frac{h^{2}}{a^{5}}B^{\alpha}=4\pi G
e_{0}^{\beta}\Xi^{\alpha}_{\beta}\nonumber\\
\end{eqnarray}
These equations can be brought to the gauge invariant form with the
use of the gauge invariant variables  defined in (25) , (31) and the
background field equations.After some manipulations, the gauge
invariant field equations are

\begin{eqnarray}
\Bigg[&-&\frac{1}{2a^{3}}\Phi'_{2}+\frac{h'+a}{a^{3}}(\Phi_{1}
+\Phi_{2})\Bigg]\delta_{i}^{\nu}+\frac{5h^{2}}{a^{2}}\partial^{\nu}\partial_{i}(\Phi_{2}-\Phi_{1})\nonumber\\
&=&4\pi G
a\Bigg(\Phi_{\pi}-2h\Phi_{q}+\Phi_{p}\Bigg)\delta_{i}^{\nu}
\end{eqnarray}

\begin{equation}
\frac{1}{2a^{3}}\partial^{\nu}\Bigg(h\Phi'_{1}-\Phi'_{2}-h\Delta\Phi_{2}\Bigg)=4\pi
G \frac{1}{a}(\Phi_{\rho}+\Phi_{q})u^{\nu}
\end{equation}

\begin{equation}
\frac{-3h}{a^{3}}\Delta\Phi_{2}+\frac{h}{a^{3}}(\Phi_{2}-\frac{h^{}}{a^{2}}\Phi_{1})=4
\pi G \bigg(\frac{-1}{a}\Phi_{\rho}\bigg)
\end{equation}

\begin{equation}
\frac{1}{2a^{3}}\partial_{a}(\Phi_{2}+h\Phi_{1})=4\pi G
a(\Phi_{\rho}+\Phi_{q})u_{a}
\end{equation}
where the first two equations are dynamical equations and the last
two will act as constraints. Note that $u^{\nu}$ and $u_{i}$ are the
4-velocity of the unperturbed background energy-momentum tensor.

\section{Dynamics of an Scalar Field as an example}
In order to find the observational consequences of the formalism
described above, we examine a case where the matter content is
described by a single scalar field. This can be applied to the
inflationary era or dark energy models. In this case, for the
background values of the energy - momentum tensor we have
\begin{equation}
\rho=\frac{1}{2}\dot{\phi}^{2}+V(\phi) \quad    ,    \quad  p=
\frac{1}{2}\dot{\phi}^{2}-V(\phi)
\end{equation}
Perturbing the scalar field as
\begin{equation}
\phi(x,t)=\phi(t)+\delta\phi(x,t)
\end{equation}
and substituting in eq. (28), we get the components of the perturbed
energy momentum tensor.

As usual we expand an  scalar fields $\chi$ into its Fourier modes
\begin{equation}
\chi(t,x)=\int
\frac{d^{3}k}{(2\pi)^{\frac{3}{2}}}\chi_{k}(t)e^{i\vec{k}.\vec{x}}
\end{equation}

\begin{figure}[htp]
\begin{center}
\includegraphics{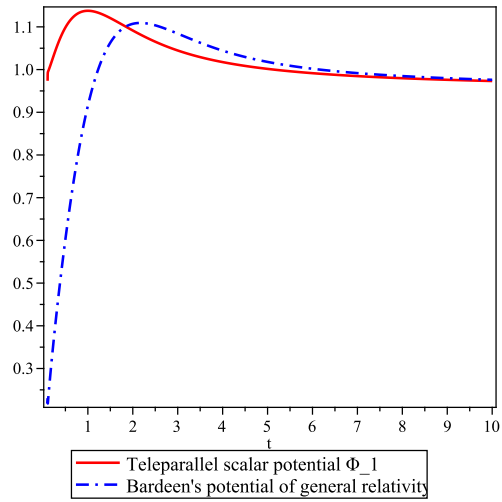} \vspace{2.5cm}
\end{center}\vspace{4cm}\caption{\small {Scalar teleparallel perturbation
potentials $\Phi_{1}$  and its general relativistic counterparts for
chaotic potential $V(\phi)=\frac{1}{2}m^{2}\phi^{2}$ . The plots are
for the wavenumber value $k=\frac{1}{1000}\,h Mpc^{-1}$. The time
axis is in Gigayears }} \vspace{1.5cm}
\end{figure}
\begin{figure}[htp]
\begin{center}
\includegraphics{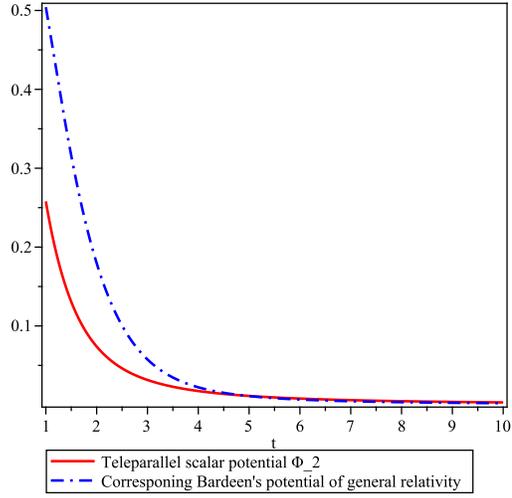} \vspace{3cm}
\end{center}
\vspace{1.5cm} \caption{\small {Scalar teleparallel perturbation
potentials $\Phi_{2}$  and its general relativistic counterparts for
the chaotic potential $V(\phi)=\frac{1}{2}m^{2}\phi^{2}$ . The plots
are for the wavenumber value $k=\frac{1}{1000}\,h Mpc^{-1}$. The
time axis is in Gigayears }}
\end{figure}
Here we are concerned with a case where there is no anisotropic
stress and no momentum perturbation. In this case we can solve
equations (39) and (40) together to find the two invariant
perturbation potentials $\Phi_{1}$ and $\Phi_{2}$ and then use one
of the constraint equations (41) or (42) to find the gauge invariant
density perturbation $\Phi_{\rho}$. The numerical analysis is done
for three types of cosmologically interesting potentials.
\\
$\bullet$ \textbf{Chaotic potential}: Figures (1) and (2) show the
time evolution of geometric potentials $\Phi_{1}$ and $\Phi_{2}$
versus the cosmic time when the scalar field potential is assumed to
be of the chaotic type \textit{i.e.}
$V(\phi)=\frac{1}{2}m^{2}\phi^{2}$. For comparison, we also plotted
the corresponding potentials deriving from the theory of general
relativity. As can be seen from the figures, the results are
different at early times (high energies) but coincide at late times.
\\
$\bullet$ \textbf{Exponential potential}: Figures (3) and (4) show
the time evolution of geometric potentials $\Phi_{1}$ and $\Phi_{2}$
versus the cosmic time for  potential in the form of
$V(\phi)=V_{0}e^{-\lambda \phi}$. Here again the results show the
same trend, however the potentials coincide at at a slightly later
time than the previous case.
\\
$\bullet$ \textbf{Power law potential}: Figures (5) and (6) show the
results for potential in the form of $V(\phi)=\lambda \phi^{4}$. In
this case the potentials for teleparallel and general relativity
theories coincide at an earlier time than the first case.
\begin{figure}[htp]
\begin{center}
\includegraphics{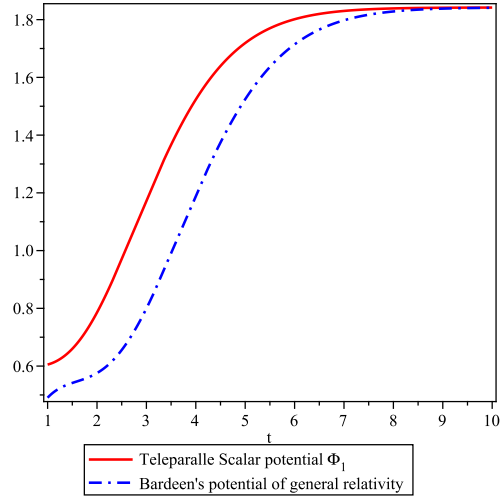} \vspace{2.5cm}
\end{center}\vspace{4cm}\caption{\small {Scalar teleparallel perturbation
potentials $\Phi_{1}$  and its general relativistic counterparts for
exponential potential $V(\phi)=V_{0}e^{-\lambda \phi}$ . The plots
are for the same value of wavenumber $k$ as in previous figures}}
\end{figure}
\begin{figure}[htp]
\begin{center}
\includegraphics{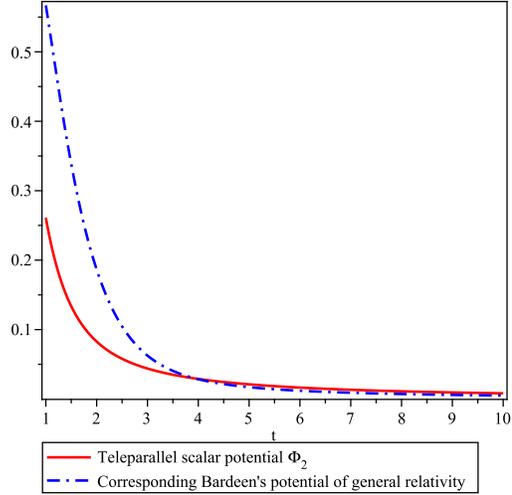} \vspace{4cm}
\end{center}
\vspace{1.5cm} \caption{\small {Scalar teleparallel perturbation
potentials $\Phi_{2}$  and its general relativistic counterparts for
exponential potential $V(\phi)=V_{0}e^{-\lambda \phi}$ and the same
value of the wavenumber $k$ as before . }}
\end{figure}
\begin{figure}[htp]
\begin{center}
\includegraphics{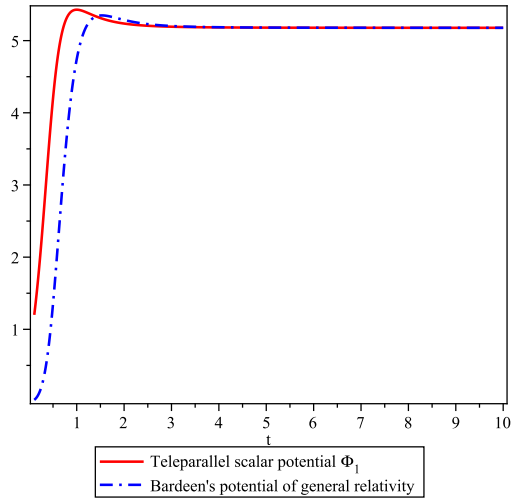} \vspace{2.5cm}
\end{center}\vspace{4cm}\caption{\small {Scalar teleparallel perturbation
potentials $\Phi_{1}$  and its general relativistic counterparts for
potential  $V(\phi)=\lambda \phi^{4}$. Other parameters are the same
as previous figures.}}
\end{figure}
\begin{figure}[htp]
\begin{center}
\includegraphics{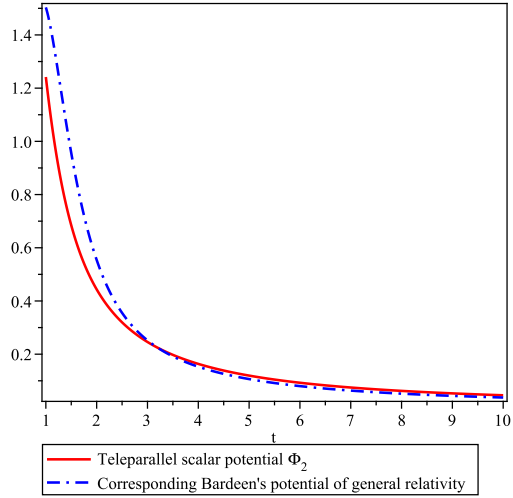} \vspace{4cm}
\end{center}
\vspace{1.5cm} \caption{\small {Scalar teleparallel perturbation
potentials $\Phi_{2}$  and its general relativistic counterparts for
potential  $V(\phi)=\lambda \phi^{4}$ .}}
\end{figure}
\newpage

\section{Conclusion}
The theory of TEGR is obtained from gauging the translational part
of the Lorentz group and after applying some restrictions on the PGT
Lagrangian will turn into a theory which is also invariant under
local Lorentz transformation. These properties has led the theory to
be dynamically equivalent to GR while they still have some
conceptual differences. This equivalence has led to a reluctance in
studying many physical issues in the context of TEGR but the special
geometric structure of this theory has some interesting features to
study. In the context of cosmological perturbations the results of
directly perturbing FRW vierbeins, are found to be different from
the results obtained from perturbing FRW metric in general
relativity. This difference stems from different transformational
properties of tetrad and metric under a gauge transformation. These
transformational properties are obtained by considering Lie
derivative of the dynamical variables of the theory. In a
teleparallel setup, one may use the gauge covariant Lie derivative
which is not equivalent to the ordinary Lie derivative used in
general relativity. Another difference comes from the fact that
tetrad perturbations only have scalar and vector modes in the
coordinate frame. As a result, the tensorial part will manifest
itself as a combination of the vector perturbations modes in the
metric. Studying cosmological issues in this context and comparing
the results by observational data, concerning all the points
mentioned above, seem to be interesting. In this paper also the
gauge invariant perturbed field equations have been found and by
means of gauge potentials. The formalism developed here can be
readily extended for use in $f(T)$ theories of gravity. However in
the $f(T$, case extreme care is needed in defining the gauge
invariant potentials as some of these models show violations of
local Lorentz invariance. In that case a gauge invariant potential
should remain unchanged under both a general coordinate
transformation \emph{and} a local lorentz rotation of the frame.

\appendix
\section{}
In this appendix we present the non-zero components of the torsion
and contorsion tensors, superpotential and the torsion scalar of the perturbed tetrad (19)\\

\textbf{Metric and inverse metric}\\
$$
\nonumber g_{00}=-a^{2}(1+2A)\quad,\quad
g_{0\alpha}=a^{2}(B_{\alpha}-C_{\alpha}) \quad,\quad
g_{\alpha\beta}=a^{2}(\delta_{\alpha\beta}+D_{\alpha\beta})\quad,\quad$$
$$ g^{00}=-\frac{1}{a^{2}}(1-2A)\quad,\quad
g^{0\alpha}=\frac{1}{a^{2}}(C^{\alpha}-B^{\alpha})\quad\quad,\quad\quad
g^{\alpha\beta}=\frac{1}{a^{2}}(\delta^{\alpha\beta}-D^{\alpha\beta})
$$\\

\textbf{Weitzenb\"{o}ck connection coefficients}\\
$$\Gamma_{~~\alpha\beta}^{\gamma}=\partial_{\beta}D_{~~\alpha}^{\gamma}\quad,\quad
\Gamma_{~~0\beta}^{\gamma}=\partial_{\beta}B^{\gamma}\quad,\quad
\Gamma_{~~\alpha0}^{\gamma}=h\delta^{\gamma}_{\alpha}+D_{~~\alpha}^{'\gamma}\quad,\quad\Gamma_{~~00}^{\gamma}=B^{'\gamma}$$
$$\Gamma_{~~\alpha\beta}^{0}=\partial_{\beta}C_{\alpha}\quad,\quad
\Gamma_{~~0\beta}^{0}=\partial_{\beta}A \quad,\quad
\Gamma_{~~\alpha0}^{0}=C'_{\alpha}\quad,\quad
\Gamma_{~~00}^{0}=h+A'$$\\
\\

\textbf{Torsion}\\
$$T_{0\alpha}^{0}=C'_{\alpha}-\partial_{\alpha}A\quad,\quad
T_{\alpha\beta}^{0}=2\partial_{\,\,[\alpha}C_{\beta]}$$
$$T_{\beta0}^{\alpha}=h\delta_{\beta}^{\alpha}-\partial_{\beta}B^{\alpha}+D_{\beta}^{'\alpha}\quad,\quad
T_{\beta\gamma}^{\alpha}=2\partial_{\,[\beta}D_{\gamma]}^{\alpha}$$\\
\textbf{Contorsion}\\
$$K_{~~0}^{\alpha0}=\frac{-1}{a^{2}}\Big[C^{'\alpha}-\partial^{\alpha}A\Big]$$
$$K_{~~0}^{\alpha\beta}=\frac{1}{a^{2}}\Big[2D^{'[\alpha\beta]}-\partial^{[\beta}B^{\alpha]}-\partial^{[\alpha}C^{\beta]}\Big]$$
$$K_{~~\beta}^{\alpha0}=\frac{1}{a^{2}}\Big[-\partial_{(\beta}B^{\alpha)}+\partial^{[\alpha}C_{\beta]}+h(1-2A)\delta_{\beta}^{\alpha}
-D_{~~\beta}^{'\alpha}\Big]$$
$$K_{~~\gamma}^{\alpha\beta}=\frac{-1}{a^{2}}\Big[\partial^{[\beta}D_{~~\gamma]}^{\alpha}
-\partial^{[\alpha}D_{~~\gamma]}^{\beta}-\partial^{[\alpha}D_{\gamma}^{~~\beta]}$$
$$+2h(\delta^{~[\alpha}_{\gamma}C^{\beta]}-\delta^{~[\alpha}_{\gamma}B^{\beta]})\Big]$$\\
\textbf{Superpotential}\\
$$S^{~~\alpha\beta}_{\gamma}=\frac{1}{2a^{2}}\Big[-\partial^{[\beta}D_{~~\gamma]}^{\alpha}+\partial^{[\alpha}D_{~~\gamma]}^{\beta}
+2\delta^{\alpha}_{\gamma}\partial^{[\beta}D_{~~\eta]}^{\eta}$$
$$-2\delta^{\beta}_{\gamma}\partial^{[\alpha}D_{~~\eta]}^{\eta}-2(\delta^{~[\alpha}_{\gamma}C^{'\beta]}-\delta^{~[\alpha}_{\gamma}\partial^{\beta]}A)$$
$$~~~~~+2h(\delta^{~[\alpha}_{\gamma}C^{\beta]}-\delta^{~[\alpha}_{\gamma}B^{\beta]})\Big]$$
$$S^{~~\alpha0}_{\gamma}=\frac{1}{2a^{2}}\Big[-\partial_{(\gamma}B^{\alpha)}+\delta^{\alpha}_{\gamma}\partial_{\eta}B^{\eta}
+\partial^{[\alpha}C_{\gamma]}+D_{~~~\gamma)}^{'(\alpha}-\delta^{\alpha}_{\gamma}D_{~~\eta}^{'\eta}$$$$-2\frac{h}{a^{2}}\delta^{\alpha}_{\gamma}
(1-2A)\Big]$$
$$S^{~~\alpha\beta}_{0}=\frac{1}{2a^{2}}\Big[D^{'[\alpha\beta]}-\partial^{[\beta}B^{\alpha]}-\partial^{[\alpha}C^{\beta]}\Big]$$
$$S^{~~\alpha0}_{0}=-\frac{1}{a^{2}}\partial^{[\alpha}D_{~~\eta]}^{\eta}-\frac{3h}{a^{2}}(C^{\alpha}-B^{\alpha})$$
\textbf{Torsion scalar}\\
$$T=-6\frac{h}{a^{2}}\partial_{\eta}B^{\eta}-\frac{h}{a^{2}}\partial^{[\eta}C_{\eta]}+6\frac{h^{2}}{a^{4}}D_{~~\eta}^{'\eta}
+6\frac{h^{2}}{a^{4}}(1-2A)$$


\begin{thebibliography}{19}
\bibitem{Baar82}
J.M. Bardeen, Phys. Rev. D22, 1882 (1980); J.M. Bardeen, P. J.
Steinhardt and M. S. Turner, Phys. Rev. D28, 679 (1983).
\bibitem{KS84}
H. Kodama and M. Sasaki, Frog. Theor. Phys. Suppl. No. 78 (1984) 1.
\bibitem{mfb92}
V.F. Mukhanov, H.A. Feldman and R.H. Brandenberger, Phys. Rept. 215,
203 (1992).
\bibitem{Durrer89}
R. Durrer, Astron. Astrophys. 208 (1989) 1.
\bibitem{Riotto}
A. Riotto, arXiv:hep-ph/0210162
\bibitem{Ein30}
A. Einstein, Math. Annal. 102, 685 (1930). For an english
translation, see A. Unzicker and T. Case, [physics/0503046v1].
\bibitem{Hehl07}
 F. W. Hehl and Y. N. Obukhov,  Annales Fond.Broglie. 32 (2007) 157-194 arXiv:gr-qc/0711.1535
\bibitem{Haya79}
K. Hayashi and T. Shirafuji, Phys. Rev. D 19, 3524 (1979); K.
Hayashi and T. Shirafuji, Phys. Rev. D 24, 3312 (1981).
\bibitem{Ald10}
R. Aldrovandi and J. G. Pereira, An Introduction to Teleparallel
Gravity, Instituto de Fisica Teorica, UNSEP, Sao Paulo, (2010).
\bibitem{TG}
R. Aldrovandi, J. G. Pereira, Teleparallel Gravity An Introduction,
Fundamental Theories of physics, Springer (2013)
\bibitem{LI}
T. P. Sotiriou, B. Li and J. D. Barrow,  Phys .Rev. D 83 , 104030
(2011)
\bibitem{FT}
R. Ferraro and F. Fiorini, Phys. Rev. D 75, 084031 (2007); N.
Tamanini and C. G. Boehmer, Phys. Rev. D 86 , 044009 (2012); Y. -P.
Wu, C. -Q. Geng, [arXiv:1110.3099]; M. E. Rodrigues, M. J. S.
Houndjo, D. Saez-Gomez, F. Rahaman,  Phys. Rev. D 86, 104059 (2012);
K. Bamba, R. Myrzakulov, S. Nojiri and S. D. Odintsov, Phys. Rev. D
85, 104036 (2012); K. Bamba, S. Capozziello, S. Nojiri, S. D.
Odintsov [arXiv:1205.3421].
\bibitem{NO1}
S. Nojiri and S. D. Odintsov, Phys. Rept. 505, 59-144,(2011),
arXiv:1011.0544 [gr-qc]
\bibitem{NO2}
S. Nojiri and S. D. Odintsov, Int.J.Geom.Meth.Mod.Phys.4,115-146
(2007), [hep-th/0601213]
\bibitem{Blag}
M. Blagojevic, Gravitation and Gauge Symmetries (IoP Publishing,
Bristol, 2002); M. Blagojevic, Three lectures on Poincare gauge
theory, SFIN A 1, 147–172 (2003) [gr-qc/ 0302040]; M. Blagojevic and
F.W. Hehl (eds.), GAUGE THEORIES OF GRAVITATION A Reader with
Commentaries Foreword by T.W.B. Kibble, FRS Imperial College Press,
London, February 2013
\bibitem{MH}
F. W. Hehl and B. Mashhoon, Phys. Lett. B 673, 279 (2009) [arXiv:
0812.1059 [gr-qc]]; F. W. Hehl and B. Mashhoon, Phys. Rev. D 79,
064028 (2009) [arXiv: 0902.0560 [gr-qc]].
\bibitem{Weitz23}
R. Weitzenb\"{o}ck, Invariance Theorie, Nordhoff, Groningen, 1923.
\bibitem{Stewart90}
J.M. Stewart, Class. Quant. Grav. 7, 1169 (1990).
\bibitem{SW}
J. M. Stewart and M. Walker, Proc. R. Soc. A (1974) 341 49-74
\bibitem{MAG}
 F. W. Hehl, J. D. McCrea, E. W. Mielke, and Y. Ne'eman, Metric affine gauge theory of gravity:
Field equations, Noether identities, world spinors, and breaking of
dilation invariance, Phys. Rept. 258, 1–171 (1995).
\bibitem{Nester1}
J.M. Nester: Gravity, torsion and gauge theory, in: Introduction to
Kaluza-Klein theories, H.C. Lee, ed. (World Scientific, Singapore
1984) pp. 83-l 15.
\bibitem{Nester2}
J.M. Nester: Lectures on gravitational gauge theory, Hsinchu School
on Gravitation, Relativity and Cosmology, Hsinchu, Taiwan, I l-13
Nov. 1989.
\end{thebibliography}
\end{document}